\documentclass[11pt,a4paper]{article}
\usepackage[hyperref]{acl2019}
\usepackage{times}
\usepackage{latexsym}

\usepackage{url}
\usepackage{graphicx}
\usepackage{amsmath}

\aclfinalcopy 

\usepackage{acronym}
\acrodef{ANN}{Artificial Neural Network}
\acrodef{ASR}{Automatic Speech Recognition}
\acrodef{ASG}{Auto Segmentation Criterion}
\acrodef{AUC}{area under the curve}
\acrodef{CCA}{Canonical Correlation Analysis}
\acrodef{CNN}{Convolutional Neural Network}
\acrodef{CTC}{Connectionist Temporal Classification}
\acrodef{DL}{Deep Learning}
\acrodef{EEG}{Electroencephalography}
\acrodef{ERP}{Event-Related Potential}
\acrodef{GAN}{Generative Adversarial Network}
\acrodef{Grad-CAM}{Gradient-weighted Class Activation Mapping}
\acrodef{GradNAP}{Gradient-adjusted Neuron Activation Profile}
\acrodef{G2P}{Grapheme-to-Phoneme}
\acrodef{CMUDict}{CMU Pronunciation Dictionary}
\acrodef{LSTM}{Long Short-Term Memory}
\acrodef{HMM}{Hidden Markov Model}
\acrodef{IPA}{International Pronunciation Alphabet}
\acrodef{LRP}{layer-wise relevance propagation}
\acrodef{MLP}{Multi-Layer Perceptron}
\acrodef{NAP}{Neuron Activation Profile}
\acrodef{NAvAI}{Normalized Averaging of Aligned Inputs}
\acrodef{NLP}{Natural Language Processing}
\acrodef{ReLU}{Rectified Linear Unit}
\acrodef{RNN}{Recurrrent Neural Network}
\acrodef{TTS}{text-to-speech}

\title{Visualizing Deep Neural Networks for Speech Recognition \\ with Learned Topographic Filter Maps}

\author{Andreas Krug \\
  University of Potsdam \\
  Research Focus Cognitive Science \\
  \texttt{ankrug@uni-potsdam.de} \\\And
  Sebastian Stober \\
  Otto von Guericke University Magdeburg \\
  Artificial Intelligence Lab \\
  \texttt{stober@ovgu.de} \\}

\date{}

\begin{document}
\maketitle
\begin{abstract}
The uninformative ordering of artificial neurons in Deep Neural Networks complicates visualizing activations in deeper layers.
This is one reason why the internal structure of such models is very unintuitive.
In neuroscience, activity of real brains can be visualized by highlighting active regions.
Inspired by those techniques, we train a convolutional speech recognition model, where filters are arranged in a 2D grid and neighboring filters are similar to each other.
We show, how those topographic filter maps visualize artificial neuron activations more intuitively.
Moreover, we investigate, whether this causes phoneme-responsive neurons to be grouped in certain regions of the topographic map.
\end{abstract}

\section{Introduction}

Improving the performance of \ac{DL} models is often achieved by increasing their complexity in number of neurons \cite{Szegedy2015}.
In this regard, their complexity becomes closer to that of real brains.
This inspires utilizing established methods from neuroscience to analyze \acp{ANN} \cite{Kriegeskorte2008}.
In our previous work, we investigated adaptations of the \ac{ERP} technique \cite{Luck2005} to analyze neural networks.
For a convolutional speech recognizer, we analyzed network responses for words \cite{Krug2017}, predicted graphemes \cite{Krug2018a} and phonemic categories \cite{Krug2018b}.
For deeper layers, we found that neuron activations are not visually interpretable, because their order within a layer is permutable.

Commonly, real brain activity is measured with \ac{EEG} and the electrode activations are visualized as a topographic map.
Such maps are top-view images of the head, which show the electrode activations at their respective location \cite{Koles1991}.
For areas between different electrode locations, the activation is interpolated.
Adapting this technique to visualize \acp{ANN} needs a topography of artificial neurons.

In this work, we adopt a regularization strategy by \citet{Kavukcuoglu2009} to learn topographic maps of filters.
The authors aimed to automatically learn locally-invariant feature descriptors for sparse coding algorithms in an unsupervised fashion.
Their idea was to arrange filters in a 2D grid and encourage similarity of neighboring coefficients.
Here, we use this strategy to learn convolutional filters in a 2D grid to visualize neuron activations as topographic maps.
We hypothesize that different regions in the 2D grid are active for particular groups of inputs.
We investigate, whether there are distinct regions in the topographic map which are active for particular phonemes.

\begin{figure*}[h]
	\centering
	\includegraphics[width=\linewidth]{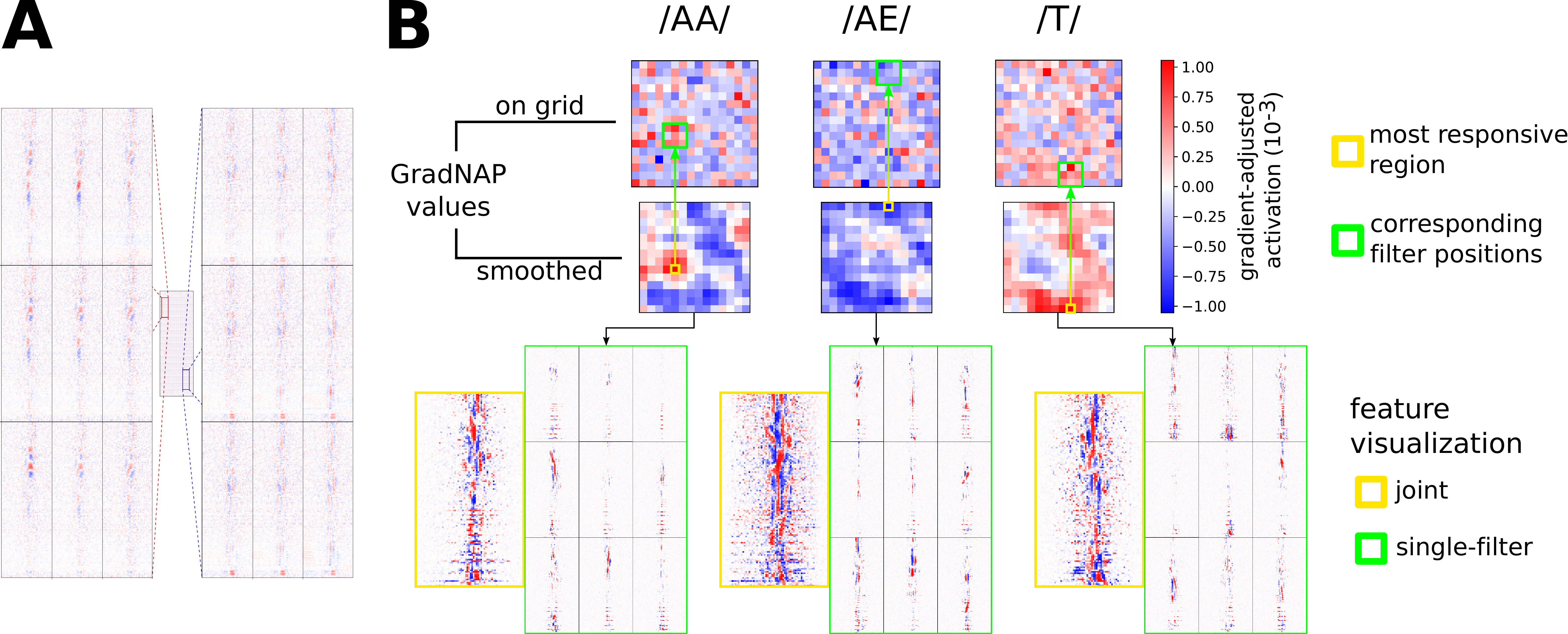}
	\caption{Close-up views on the regularized filter space in the first layer (A). Network responses to exemplary phonemes in the second layer (B top) and feature visualization for the most responsive region (B bottom).}
	\label{fig:results}
\end{figure*}

\section{Methods}
\subsection{Learning topographic filter maps}
We learn topographic filter maps by constraining the optimization.
First, a grid is defined for every layer.
We arrange layers with 256 filters in 16$\times$16 grids and 2048-filter layers in 64$\times$32 grids.
Filters are encouraged to be similar within a n$\times$n-neighborhood.
For each neighborhood, which a filter is included in, the similarity to the center filter is computed.
Those similarities are weighted by the reciprocal Euclidean distance to the center position, while the weights are normalized to sum up to 1 per neighborhood.
The loss function constraint is the sum of weighted similarities over all filters and all neighborhoods.
Here, we minimize Cosine similarity in a 3$\times$3 neighborhood.
The Cosine similarity between two vectors $A$ and $B$ is defined as $d=\left(A\cdot B\right)/\left(||A||*||B||\right)$.
As this constraint mainly affects the order of the neurons, it negligibly interferes with model performance.
Figure \ref{fig:results}A shows two exemplary neighborhoods in a 16$\times$16 topographic filter map.

\subsection{Determining group-specific network responses}
To characterize how the network responds to particular groups of inputs, we use \acp{GradNAP} (under review).
\acp{GradNAP} are an extension of our previously described \acp{NAP} \cite{Krug2018b}.
This method is inspired by the \ac{ERP} technique.
The idea is to average neuron activations for all inputs corresponding to the same group.
For our speech recognizer, we averaged activations over particular phonemes or graphemes.
Due to normalizing the average activations by subtracting baseline activations, \ac{GradNAP} values can be negative or positive.
	
\section{Results \& Discussion}
Figure~\ref{fig:results}B shows results of our method for 3 exemplary phonemes.
The square plots (top half) show time-averaged \ac{GradNAP} values on the learned topographic map.
Plotting the values directly (``on~grid'') is hard to visually interpret, because the values do not show smooth transitions.
To achieve a visually appealing topographic map, we apply non-strided average pooling in 3$\times$3 windows (``smoothed'').
In the resulting map, we locate the maximum value to identify the 3$\times$3-region of highest phoneme-responsiveness.
For the identified regions, we compute optimal inputs \cite{Yosinski2015} for each filter in the region separately and jointly for the responsive neighborhood.
The optimal inputs are shown in the bottom row of Figure~\ref{fig:results}B, ``single-filter'' and ``joint'', respectively.
We observed that only considering the single most responsive region does not reveal phoneme-typical patterns.
This indicates that the representations are still more distributed on the grid.
For example, many regions of the grid are strongly activated for phoneme \texttt{/T/}.
The most responsive region is therefore likely missing some parts of the distributed representation.

\section{Conclusion}
Topographic filter maps are a promising way of using well-established methods from neuroscience to visualize Deep Neural Networks.
The learned ordering of the neurons allows to show activations in a way which is more intuitive for a person.
However, the current similarity constraint is not encouraging enough representational sparsity.
Features are still distributed between too many regions in the topographic map.
Therefore, optimizing inputs for particular regions in the grid does not yield interpretable patterns.
In future work, we will investigate more regularization strategies.
This will include adapting the constraint to regularize activations instead of filter weights and incorporating global similarity penalties.

\section*{Acknowledgments}

This research has been funded by the Federal Ministry of Education and Research of Germany (BMBF) and supported by the donation of a GeForce GTX TitanX graphics card from the NVIDIA Corporation.

\bibliography{acl2019}
\bibliographystyle{acl_natbib}

\appendix

\end{document}